\documentclass[paper]{sigirforum}
\usepackage{amsthm}
\usepackage[inline]{enumitem}
\newtheorem{definition}{Definition}

\usepackage[dvipsnames]{xcolor}
\newcommand{\heading}[1]{\bigskip\noindent\textbf{#1}}

\def\pubissue{Vol. 56 No. 2 December 2022}

\begin{document}
\title{Offline Evaluation for Reinforcement Learning-based Recommendation: \\ A Critical Issue and Some Alternatives}

\authors{
\author[\hspace{-0.6cm}romain.deffayet@naverlabs.com]{\hspace{-0.6cm}Romain Deffayet}{\hspace{-0.6cm}Naver Labs Europe \& University of Amsterdam}{\hspace{-0.6cm}France / The Netherlands}
\and
\author[\hspace{0.6cm}thibaut.thonet@naverlabs.com]{\hspace{0.6cm}Thibaut Thonet}{\hspace{0.6cm}Naver Labs Europe}{\hspace{0.6cm}France}
\and
\author[jean-michel.renders@naverlabs.com]{Jean-Michel Renders}{Naver Labs Europe}{France}
\and
\author[\hspace{2cm}m.derijke@uva.nl]{\hspace{1.9cm}Maarten de Rijke}{\hspace{1.9cm}University of Amsterdam}{\hspace{1.9cm}The Netherlands}
}

\maketitle 

\begin{abstract}
In this paper, we argue that the paradigm commonly adopted for offline evaluation of sequential recommender systems is unsuitable for evaluating reinforcement learning-based recommenders. We find that most of the existing offline evaluation practices for reinforcement learning-based recommendation are based on a \emph{next-item prediction} protocol, and detail three shortcomings of such an evaluation protocol. Notably, it cannot reflect the potential benefits that reinforcement learning (RL) is expected to bring while it hides critical deficiencies of certain offline RL agents. Our suggestions for alternative ways to evaluate RL-based recommender systems aim to shed light on the existing possibilities and inspire future research on reliable evaluation protocols.
\end{abstract}

\section{Introduction}
\label{intro}

Recommender systems play a major role in defining internet users' experience due to their ubiquitous presence on, e.g., content providing and e-commerce platforms. Correct and careful evaluation of recommender systems is therefore critical as it directly impacts business metrics as well as user satisfaction -- and sometimes even society as a whole.

While recommendation accuracy (i.e., recommending relevant items) is often taken to be the main indicator of performance, the literature on recommender systems highlights the importance of additional criteria. 
Beyond-accuracy goals include, e.g., diversity, novelty or serendipity, fairness, and user experience in general \citep{accurate-not-enough}. Such criteria sometimes cannot be enforced in one-shot recommendation (i.e., in a single interaction between the user and the recommender system) but they may require that we consider the longer-term experience. These concerns have motivated researchers and practitioners alike to acknowledge the sequential nature of many recommendation engines, and to seek to optimize over whole sequences instead of one-shot predictions~\citep{seq-aware-RS}. 

Reinforcement learning (RL) formulates this problem as a Markov decision process (MDP), in which we wish to select appropriate actions (i.e., item recommendations) in order to maximize the sum of rewards (e.g., clicks, purchases, etc.) along the full sequence of user interactions with the recommender system. RL is a natural fit for this problem because the underlying MDP is able to model the long-term influence of recommendations on the user.
Note that in recommendation scenarios, online exploration is often impossible, so the policy must be trained from a fixed dataset of interactions, i.e., by offline RL. While sequence optimization with offline RL is not expected to entirely fulfill all the desired beyond-accuracy criteria highlighted in the literature, it holds the promise of making some of the desired properties naturally emerge as a result of whole-sequence optimization. Indeed, one can expect that, given an appropriate reward function, policies that are effective over the entire span of the user's experience require some of these desired properties: diversity, novelty, etc. Because these auxiliary metrics are embedded into the sequence's cumulative reward, whole-sequence optimization with RL can be seen as a way to bridge the gap between offline and online performance.

In this paper, we argue that the progress supposedly achieved in sequential recommendation, thanks to RL, lacks \emph{ecological validity}~\citep{andrade-2018-internal}: the trained agents are likely not to generalize to real-world scenarios, because of certain shortcomings in the current evaluation practices. Namely, RL-based recommender systems are often evaluated in an offline fashion, following a traditional one-shot accuracy-oriented protocol that cannot capture the potential benefits introduced by the use of RL algorithms. We refer to this evaluation protocol as \textit{next-item prediction} (NIP). More critically, we highlight that the specifics of this protocol are likely to hide the deficiencies of recommender systems trained by offline RL. Briefly, we argue that \emph{with the most commonly employed evaluation practices, we cannot verify that the RL algorithm correctly optimizes the very metric it is designed to optimize}, i.e., expected cumulative reward. We worry that instead of bridging the gap between offline and online performance, it only widens it. We then provide suggestions towards a sound evaluation methodology for RL-based recommendation in order to help practitioners and researchers avoid common pitfalls and to inspire future research on this important topic.

After contrasting our criticism with that formulated by previous studies in Section \ref{related}, we provide in Section \ref{whatisit} a definition of the \textit{next-item prediction} evaluation protocol along with an overview of its use in sequential recommendation with RL. Section \ref{whatswrong} dives into the three major issues of the NIP protocol, and their implications for the evaluation of RL-based recommender systems. Finally, we formulate our suggestions towards a sound evaluation methodology in RL-based recommendation in Section \ref{whattodo}.


\section{Related studies}
\label{related}

Deficiencies in recommender systems evaluation have been a long-standing problem in the recommendation literature. In this section we review previous studies that discuss this topic.

Firstly, as we recalled in the introduction, \citet{accurate-not-enough, beyond-matrix-completion} have highlighted the need for recommender systems that go beyond accuracy of the proposed item, i.e., which do not only consider recommendation as a matrix completion problem. This is motivated by an observed gap between offline and online performance, sometimes rendering any conclusions drawn from offline evaluation obsolete \citep{swissinfo, netflix, jeunen-offline-eval}. 

Secondly, pitfalls of recommender system evaluation~-- including the next-item prediction protocol for offline evaluation that we focus on in this study~-- have been extensively discussed in the past: \citet{common-pitfalls, jeunen-offline-eval, data-leakage, recommender-systems-crisis, fresh-look-RS, revisiting-study} highlighted multiple issues resulting from data leakage and other dataset construction fallacies, which can lead to counter-intuitive statements. The presence of selection bias in the data used for evaluating recommender systems from implicit feedback has also been identified as a major source of inaccuracies \citep{netflix, beyond-matrix-completion, common-pitfalls, jeunen-offline-eval}. In addition, and more specifically to the next-item prediction protocol, \citet{sampled-metrics, revisiting-study} have shown that sampling negative items at inference time in order to ease the computation of ranking metrics leads to drawing incorrect conclusions on the recommendation performance.

Finally, many studies reaffirm the importance of appropriate baseline selection in order to ensure that progress has been made, and have shown that certain claims do not hold against properly tuned baselines \citep{session-based-eval, much-progress, tuning-baselines, benchmarking-reco, revisiting-study}.

The argument we formulate in this paper is specific to RL-based recommendation and while it has, to the best of our knowledge, never been expressed, it is not incompatible with the issues listed in this section. It is rather to be considered as an additional caveat when evaluating RL-based recommender systems.

\section{Next-item prediction in RL-based recommendation}
\label{whatisit}

We propose an (informal) definition of \textit{next-item prediction} that encompasses the offline evaluation protocols of many sequential recommendation studies, and that we consider to be problematic when used to evaluate RL-based recommender systems:

\begin{definition}
\label{def}
\rm
\emph{Next-item prediction} (NIP) is an offline evaluation protocol for sequential item recommendation from real user feedback. The task is to ensure that the next interacted item is among the top items ranked by the model, given the sequence of past interactions. Model performance is measured according to ranking metrics (e.g., hit rate, recall, NDCG, etc).
\end{definition}

\noindent We propose this definition because it is representative of the evaluation setup adopted in many sequential recommendation studies, e.g., GRU4REC~\citep{gru4rec}, and also encompasses several variants. In particular, the choice of ``next interacted item'' can vary depending on the dataset and task at hand: the next clicked item in content recommendation (e.g., Last.fm \citep{lastfm}), the next purchased product in product recommendation (e.g., RecSys Challenge 2015~\citep{challenge15} or RetailRocket~\citep{retailrocket}), the next highly rated movie in movie recommendation (e.g., MovieLens~\citep{movielens}), the next basket in grocery shopping~\citep{instacart}, etc. 

\begin{figure}[t]
    \centering
    \includegraphics[width=0.6\textwidth]{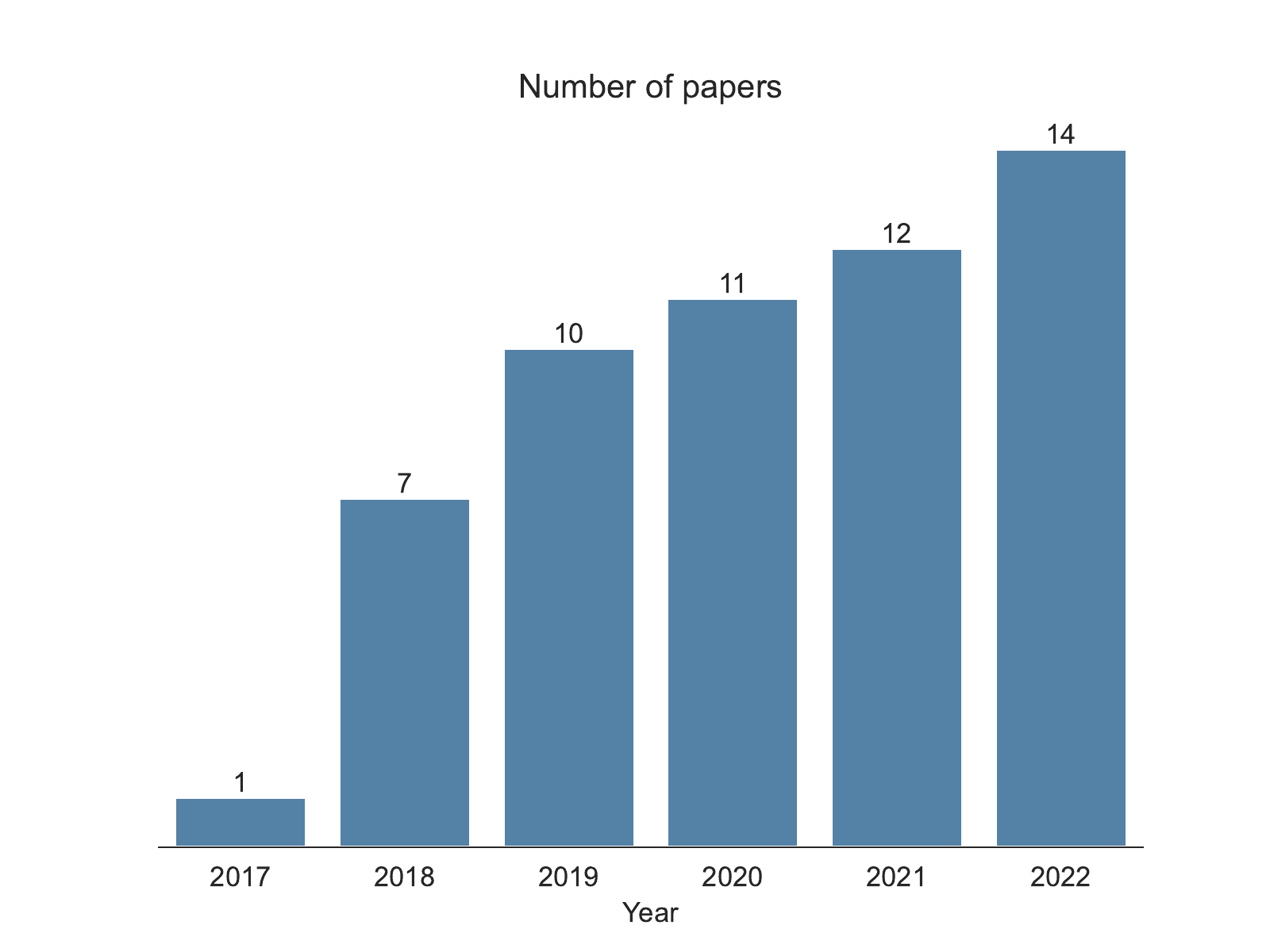}
    \caption{Evolution of the number of RL-based recommendation papers published in major RecSys and IR conferences between 2017 and 2022.}
    \label{fig:rl4rec_papers}
\end{figure}

\heading{How prevalent is it in RL-based recommendation?}
RL-based recommendation (RL4REC) has become increasingly popular in recent years: we counted $55$ papers about RL4REC in the proceedings of major information retrieval and recommender systems (or related) conferences between January 2017 and October 2022. To obtain this result, we queried ``reinforcement learning recommendation'' and ``reinforcement learning recommender'' on DBLP\footnote{\url{https://dblp.org/}} and included papers published at AAAI, CIKM, ICDM, IJCAI, KDD, RecSys, SIGIR, WSDM or WWW. Figure \ref{fig:rl4rec_papers} shows the increasing trend in published RL4REC papers.
Out of the 55 papers retrieved from DBLP, we identified 39 papers that address sequential item recommendation using RL-based approaches. Other tasks irrelevant to our argument included conversational recommendation or explainable recommendations, so we ignore papers related to these topics in this study. Among the 39 relevant articles, we found 24 papers performing a form of offline evaluation, including 22 papers that followed the NIP protocol from Definition~\ref{def}. The 15 other papers exclusively rely on online evaluation, either in production using an industrial recommendation platform or based on a simulator. \emph{The NIP protocol is therefore by far the most commonly adopted type of offline evaluation.}

\section{Three shortcomings of NIP}
\label{whatswrong}

Before engaging with the explanation of the issues with next-item prediction, we would like to recall the benefits promised by the use of RL algorithms:
\begin{itemize}[nosep]
    \item RL aims to optimize long-term outcomes resulting from a sequence of decisions. This requires accounting for the effect of the recommender on the user. RL-based methods are able to optimize whole-sequences by assigning the credit for observed rewards to individual actions, thereby preventing costly search throughout the combinatorial space of action sequences.
    \item RL algorithms learn in a self-supervised manner, by maximizing scalar rewards. Doing so allows them to recover open-ended solutions and generate novel policies. However, training the agent in an offline fashion also comes with the risk of deriving policies with inaccurate estimation of their expected return.
\end{itemize}

\noindent%
In the following, we list three major shortcomings of the NIP protocol for evaluating offline RL agents, and explain how they harm the ecological validity of the claims derived from this evaluation protocol. 

\subsection{A myopic evaluation} Evaluating an offline RL-based recommender system using Definition \ref{def} only accounts for short-term rewards and ignores the causal effect of the recommendations on the user. Indeed, an important motivation to design RL algorithms is to maximize the return (i.e., sum of rewards) along full trajectories, as opposed to bandit algorithms that aim to maximize the average reward at each timestep. When the actions (i.e., recommendations) cause the environment (i.e., user) to change its state, RL algorithms still have convergence guarantees, while the environment appears as non-stationary to bandit algorithms that fail to find the optimal policy both in theory and in practice. But the \textit{next-item prediction} evaluation protocol only requires short-term thinking as it rewards one-shot prediction of the next interacted item~-- this is due to the offline, static nature of the evaluation that overlooks the causal impact of the recommendation policy of interest over subsequent interactions. This argument has been formulated by \citet{dongyoon}, who also empirically verified that greedy, myopic agents achieve similar or better performance on the NIP protocol than long-term-aware RL agents on standard recommendation datasets. \citet{seq-aware-RS} also warned about the limits of the NIP evaluation protocol in sequential recommendation when not only immediate satisfaction but also diversity or user guidance in content discovery is desired.

However, in contrast to \citet{dongyoon}, we additionally argue that the inclusion of delayed rewards such as dwell-time in content recommendation or lifetime value in product recommendation would not be sufficient to solve this issue. Indeed, the long-term outcomes encoded in the delayed reward (e.g., was the product satisfactory over its whole lifetime?) can be orthogonal to the long-term outcomes encoded in the sum of rewards along the trajectory (e.g., was the trajectory diverse enough to avoid boring out the user?). While the former clearly seem to be important in order to obtain useful and enjoyable recommender systems, the latter are the ones that are modeled by the Markov decision process underlying the RL agent. Consequently, if we include delayed rewards but ignore the long-term outcomes induced by the sequential decision-making process, we still cannot observe the benefits brought by RL training from the NIP protocol. Note that these two types of long-term outcomes are not incompatible and we recommend using a reward function that is as close as possible to the user's needs and satisfaction, including delayed outcomes.

\subsection{A suboptimal target} As explained in Section \ref{whatisit}, in datasets commonly employed for next-item prediction, we observe the rewards (e.g., clicks, purchases) only on the items that the user interacted with. This incurs a selection bias in the evaluation protocol, caused by the application of a particular treatment to the user. This treatment can take the form of a logging policy or a mixture of logging policies when data is gathered from organic interactions on recommendation platforms, or the implicit effect of exogenous factors when the observed data is the result of active user feedback, e.g., voluntary movie reviews or product search. We refer to the latter kind of bias as an implicit logging policy for simplicity. Note that another source of sub-optimality of the interacted items is that user choice may also be shortsighted or reluctant to novelty, even though acting so may lead to a less enjoyable experience overall.

By considering the fact that selecting the interacted item is a binary target, instead of a scalar reward to be maximized, the NIP evaluation incentivizes researchers and practitioners to build policies that are close to the (implicit) logging policy, at the expense of choosing optimal actions. It is a close-ended task of policy matching while RL allows for open-ended outcomes, i.e., generating novel policies achieving high return. There exists simpler methods to replicate the policy which generated the data, e.g., imitation learning \citep{imitation-learning}, and the reward maximization objective of RL is likely to deteriorate the results on this evaluation by selecting items that are different from the interacted item but incurring higher returns.
Consequently, NIP will discard performant policies and encourage policies similar to the logging policy, even when the sequences in the dataset were highly suboptimal. Considering stronger signals such as purchases or high ratings mitigates this issue, but the selection bias that users were exposed to during data collection implies that some highly rewarding items are likely discarded.

\subsection{Risky deployment} 
\label{risky-deployment}
The two previous points that we have formulated indicate that the \textit{next-item prediction} evaluation cannot reflect the potential benefits brought by offline RL-based recommender systems. The third problematic aspect that we discuss shows that next-item prediction may also hide critical deficiencies of offline RL agents.

Even though in the evaluation protocol of Definition \ref{def} we account for the position of the next interacted item in the model predictions, through the use of ranking metrics, the recommender system will only select its most preferred item (or top-$k$ most preferred items in slate recommendation) when used in production, while none of the other items will be shown to the user. It therefore seems crucial to ensure that the top item is satisfactory, regardless of the full ranking. This is unfortunately not possible with a fixed dataset where only one or a few items have been shown to the considered user. A tacit assumption of NIP is that higher ranking metrics correlate with a top item causing high return. However, a gap between offline and online results has been identified in previous studies \citep{swissinfo, netflix}. More importantly, it has been shown that even under the strong assumption that the $Q$-value associated to every action (i.e., item recommendation) can be correctly estimated in expectation (i.e, no bias), there can be an overestimation of the predicted offline reward with respect to the actual online reward, because the selected item is more likely to be one of those with an overestimated $Q$-value \citep{optimizers-curse}. This phenomenon is called the \emph{optimizer's curse}, and while its practical impact in certain cases can be limited, we argue that it can critically affect RL algorithms.
Indeed, a particular set of conditions has been identified to cause a catastrophic impact of the optimizer's curse and is often called the \textit{deadly triad} \citep{deadly-triad, sutton-barto}. It can be observed with most RL algorithms and occurs when \begin{enumerate*}[label=(\roman*)]
    \item the value estimate at one state is used to update the value estimate at the previous state, 
    \item function approximation is used to build the estimate of the value function, and 
    \item the RL agent is trained in an off-policy fashion.
\end{enumerate*}
\vspace{-0.25cm}

Under such conditions, small overestimations of the value function on out-of-distribution actions can be amplified and propagated to neighboring states and actions, potentially leading to divergence of the value function. In that case, while the model predicts high $Q$-values for its policy, the observed return after deployment can be arbitrarily bad. The highly damaging effect of the deadly triad has been observed in multiple scenarios and motivated the emergence of extensive research on offline reinforcement learning \citep{deadly-triad, bottlenecks-dqn, d4rl, offline-tutorial, onestep, iql}.
Unfortunately, this harmful phenomenon cannot be detected in the standard \textit{next-item prediction} evaluation of Definition \ref{def}: while the interacted item may rightfully be ranked high by the model, it is likely that at least one out-of-distribution item is drastically overestimated and preferred by the model. Since this item will be the one selected by the model, we may observe an unpredicted catastrophic failure at deployment time. Even worse, this probability of failure tends to increase with the size of the action-space \citep{merlion}, which can be enormous in certain recommendation scenarios.

\subsection{Upshot} 
The three shortcomings we presented in this section render offline evaluation using the NIP protocol of RL-based recommender systems unreliable. They effectively widen the gap between offline and online metrics, where RL algorithms were actually supposed to bridge this gap. In the next section, we suggest potential solutions to address this issue.

\section{Some alternatives to NIP}
\label{whattodo}

The limitations of NIP make offline evaluation of RL-based recommender systems difficult. We detail below some partial solutions to this problem and discuss their limitations and remaining open questions.

\subsection{Online evaluation in recommendation platforms} The most obvious counter-measure to the issues raised above is to evaluate recommender systems online when possible, directly on the metrics we care about. This is usually done by deploying the policies on an actual recommendation platform. However, it is obvious that not all researchers and practitioners have access to an operational industrial platform, and online evaluation itself may include other forms of biases, e.g., through the inclusion of business rules in recommendations. Online evaluation clearly circumvents the three issues we highlighted in the previous section, but since the focus of this paper is on offline evaluation, we will not further detail it.

\subsection{Counterfactual off-policy evaluation} There is a large body of work on off-policy evaluation (OPE) in information retrieval, often based on techniques such as inverse propensity scoring \citep{crm-joachims, ultr-joachims}, where a propensity weight is applied to rescale the observed rewards and returns. Although OPE has mostly been tackled for the one-shot bandit problem, some studies address OPE of RL policies both in the RL community \citep{DOPE} and in the IR community \citep{topk-reinforce}, and more recently a library for off-policy evaluation of RL algorithms in IR has been proposed in \citep{ofrl}.

Counterfactual methods for off-policy evaluation are attractive in that they can provide unbiasedness guarantees under mild assumptions. However, we want to stress three (known) deficiencies of these methods:
\begin{enumerate*}[label=(\roman*)]
    \item IPS suffers from a notoriously high variance which becomes exponentially higher when applied on sequences, because of the product of inverse propensity weights \citep{per-decision-IS}; 
    \item in non-tabular settings (i.e., when one can generalize the predictions from a state-action pair to another, for example with continuous spaces), generalization capabilities must implicitly or explicitly be assumed when the logging policy is not known, in order to compute the propensity \citep{is-estimated-behavior}; and
    \item when we train RL algorithms in an offline manner, the error of the off-policy training and of the off-policy evaluation are likely correlated, which means that counterfactual OPE may still be biased and wrongly choose certain methods above others. An extreme example of the latter occurs if we train and evaluate a policy-gradient recommender with the same propensity weights, which makes the agent appear as optimal regardless of its true performance. While using an ensemble of estimators might mitigate this issue, it remains unclear how to fully alleviate this issue. Counterfactual OPE circumvents all three shortcomings highlighted in the previous section in theory, but as we have seen it comes with its own shortcomings which may make it unreliable in certain practical settings.
\end{enumerate*} 

\subsection{Simulator-based evaluation} Simulators have proved useful to assess progress in other domains, such as robotics, games or industrial applications \citep{d4rl, rl-unplugged, neoRL}. While the interaction with a recommender system is arguably one of the hardest problems to simulate because of the complexity and apparent stochasticity of human behavior, the true value of simulators lies in their ability to observe how recommenders react under a chosen set of assumptions on user behavior. Additionally, by allowing the researcher to access otherwise unobservable metrics, they can enlighten us on the inner workings of the systems we build.

Many studies proposed to build semi-synthetic simulators, where the synthetic part is as limited as possible in order to adhere to real-world scenarios. This can for instance be done by using real item embeddings \citep{taobao} or by extending the implicit feedback to unseen data, with debiasing in the missing-not-at-random case \citep{sofa}. Moreover, it is possible to assess the generalizability of a method by benchmarking it against a wide range of simulated configurations, so as to mitigate the influence of simulator design on the results. Regardless of the chosen setup, one should ensure that the simulator exhibits the characteristics we wish to model, most notably long-term influence of the recommender system on the user.

Simulators are not sensitive to the three issues of the NIP protocol, but their ecological validity may clearly be limited. On top of building simulators from real data, some approaches aim to bridge the gap between simulation and reality, for example with domain randomization \citep{domain-randomization, openai-rubiks-cube}.

\subsection{Intermediate evaluation} By intermediate evaluation, we refer to the offline evaluation of models, simulators or propensities that are used as building blocks in the final recommendation model \citep{sofa, cm-shift}. In certain cases, it may be easier to evaluate these intermediate models than the final model, for example when they can be evaluated thanks to the availability of human annotations, e.g., of item relevance. By breaking down the evaluation protocol into several components, we can isolate and reduce the sources of bias. For instance, in top-$k$ recommendation for cumulative click maximization, if the click model is correctly estimated, i.e., the relevance and propensity scores are correct, then only state dynamics (i.e., how a user changes in response to a recommendation) are left as a source of uncertainty.

Doing so mitigates the risks associated with deploying RL agents, but does not suppress them. Moreover, we want to stress that offline RL agents will likely use the intermediate models outside of their training distribution in order to perform policy evaluation, and therefore may exploit inaccuracies in these high uncertainty regions if no proper countermeasure is applied \citep{cm-shift}.

\subsection{Uncertainty-aware evaluation} While it may not be feasible to accurately evaluate the final performance of an RL policy in a purely offline fashion, we argue that quantifying its performance at different levels of uncertainty can help assess the risks of deployment. Indeed, the value overestimation issue highlighted in the previous section results from the high uncertainty on out-of-distribution state-action pairs. We can constrain the RL algorithm to recover safe policies, that stay within the distribution of the logging policy, or allow exploration in order to find potentially high-return policies, at the cost of increasing uncertainty \citep{onestep}. By quantifying the match between the support of the logging policy and that of the target policy, we can assess the risk induced by the deployment of the target policy. In particular, if we restrict the set of available actions to those considered ``in-support'', we can get an accurate estimate of the performance of the policy on those actions. Indeed, uncertainty is low inside the support of the logging policy, and it is anyway possible to evaluate the quality of the $Q$-value prediction on a held-out test set of the offline dataset as in, e.g., \citep{rl-ltv}. A safe policy achieving high in-support expected return would constitute a reliable improvement, while an unsafe policy not even achieving good in-support expected return can probably be discarded. This type of evaluation needs a proper definition of in-support and out-of-support, e.g., as in \citep{bcq, merlion}, which is not trivial in the non-tabular setting and requires assuming a certain degree of tolerance to uncertainty, but \citet{workflow-offline-rl} show that it is possible to adjust this tolerance based on the training curves of certain offline RL algorithms.

This type of evaluation focuses on characterizing and mitigating the risks induced by the third issue we raise in Section \ref{risky-deployment}, while potentially allowing us to detect the benefits brought by RL training. The main open question lies in the ability to properly define distance measures between the support of the logging and target policy.

\section{Conclusion}

In this study, we highlighted that the most commonly employed protocol for the offline evaluation of RL-based recommender systems is in fact unsuitable, because it cannot reflect the benefits that RL supposedly brings compared to more traditional approaches and because it may hide critical deficiencies of offline RL agents that can lead to catastrophic deployment. These shortcomings can be summarized as follows: \begin{enumerate*}[label=(\roman*)]
    \item a myopic protocol aimed only at measuring short\-term accuracy,
    \item a close-ended, suboptimal recommendation target, and
    \item sensitivity to the optimizer's curse.
    \end{enumerate*}

As of now, there exists no truly satisfactory solution to the problem of evaluating RL policies in an entirely offline fashion. Yet, several proxies for online performance can be used to bridge the gap between offline metrics and online performance. Finding appropriate offline evaluation protocols is still an active research area in the offline RL literature, and we urge the sequential recommendation community to join the effort and develop protocols suitable for the recommendation scenario.
Additionally, acknowledging the presence of uncertainty in the deployment of RL-based recommender systems paves the way towards solutions that are robust or resilient to such uncertainty. For instance, \citet{genspec} propose a criterion for fallback to a safer policy when out-of-distribution (although in a different context, i.e., counterfactual learning to rank), and \citet{adaptive-offline-rl, back-to-manifold} propose adaptive offline RL policies that are able to recover from stepping in uncertain states during deployment by branching back to supported states. We hope that future research in recommender systems will put stronger emphasis on these aspects and reduce the gap between offline and online performance.

\bibliography{sigirforum}

\begin{thebibliography}{54}
\providecommand{\natexlab}[1]{#1}
\providecommand{\url}[1]{\texttt{#1}}
\expandafter\ifx\csname urlstyle\endcsname\relax
  \providecommand{\doi}[1]{doi: #1}\else
  \providecommand{\doi}{doi: \begingroup \urlstyle{rm}\Url}\fi

\bibitem[Andrade(2018)]{andrade-2018-internal}
Chittaranjan Andrade.
\newblock Internal, external, and ecological validity in research design,
  conduct, and evaluation.
\newblock \emph{Indian Journal of Psychological Medicine}, 40:\penalty0
  498--499, 2018.

\bibitem[Ben-Shimon et~al.(2015)Ben-Shimon, Friedmann, Tsikinovsky, Hörle,
  Rokach, and Shapira]{challenge15}
David Ben-Shimon, Michael Friedmann, Alexander Tsikinovsky, Johannes Hörle,
  Lior Rokach, and Bracha Shapira.
\newblock Recsys challenge 2015, 2015.
\newblock URL \url{https://recsys.acm.org/recsys15/challenge/}.

\bibitem[Brandfonbrener et~al.(2021)Brandfonbrener, Whitney, Ranganath, and
  Bruna]{onestep}
David Brandfonbrener, Will Whitney, Rajesh Ranganath, and Joan Bruna.
\newblock Offline rl without off-policy evaluation.
\newblock In \emph{NeurIPS}, pages 4933--4946, 2021.

\bibitem[Chen et~al.(2017)Chen, Chung, Huang, and Tsui]{common-pitfalls}
Hung-Hsuan Chen, Chu-An Chung, Hsin-Chien Huang, and Wen Tsui.
\newblock Common pitfalls in training and evaluating recommender systems.
\newblock \emph{ACM SIGKDD Explorations Newsletter}, 19\penalty0 (1):\penalty0
  37–45, sep 2017.

\bibitem[Chen et~al.(2019)Chen, Beutel, Covington, Jain, Belletti, and
  Chi]{topk-reinforce}
Minmin Chen, Alex Beutel, Paul Covington, Sagar Jain, Francois Belletti, and
  Ed~H. Chi.
\newblock Top-k off-policy correction for a reinforce recommender system.
\newblock In \emph{WSDM}, page 456–464, 2019.

\bibitem[Cremonesi and Jannach(2021)]{recommender-systems-crisis}
Paolo Cremonesi and Dietmar Jannach.
\newblock Progress in recommender systems research: {Crisis}? {What} crisis?
\newblock \emph{AI Magazine}, 42\penalty0 (3):\penalty0 43--54, Nov. 2021.

\bibitem[Deffayet et~al.(2022)Deffayet, Renders, and de~Rijke]{cm-shift}
Romain Deffayet, Jean-Michel Renders, and Maarten de~Rijke.
\newblock Evaluating the robustness of click models to policy distributional
  shift.
\newblock \emph{ACM Trans. Inf. Syst.}, oct 2022.

\bibitem[Ferrari~Dacrema et~al.(2019)Ferrari~Dacrema, Cremonesi, and
  Jannach]{much-progress}
Maurizio Ferrari~Dacrema, Paolo Cremonesi, and Dietmar Jannach.
\newblock Are we really making much progress? {A} worrying analysis of recent
  neural recommendation approaches.
\newblock In \emph{RecSys}, page 101–109, 2019.

\bibitem[Fu et~al.(2019)Fu, Kumar, Soh, and Levine]{bottlenecks-dqn}
Justin Fu, Aviral Kumar, Matthew Soh, and Sergey Levine.
\newblock Diagnosing bottlenecks in deep {Q}-learning algorithms.
\newblock In \emph{ICML}, pages 2021--2030, 2019.

\bibitem[Fu et~al.(2020)Fu, Kumar, Nachum, Tucker, and Levine]{d4rl}
Justin Fu, Aviral Kumar, Ofir Nachum, George Tucker, and Sergey Levine.
\newblock {D4RL}: Datasets for deep data-driven reinforcement learning.
\newblock \emph{arXiv:2004.07219}, 2020.

\bibitem[Fu et~al.(2021)Fu, Norouzi, Nachum, Tucker, Wang, Novikov, Yang,
  Zhang, Chen, Kumar, Paduraru, Levine, and Paine]{DOPE}
Justin Fu, Mohammad Norouzi, Ofir Nachum, George Tucker, Ziyu Wang, Alexander
  Novikov, Mengjiao Yang, Michael~R. Zhang, Yutian Chen, Aviral Kumar, Cosmin
  Paduraru, Sergey Levine, and Tom~Le Paine.
\newblock Benchmarks for deep off-policy evaluation.
\newblock In \emph{ICLR}, 2021.

\bibitem[Fujimoto et~al.(2019)Fujimoto, Meger, and Precup]{bcq}
Scott Fujimoto, David Meger, and Doina Precup.
\newblock Off-policy deep reinforcement learning without exploration.
\newblock In \emph{ICML}, 2019.

\bibitem[Garcin et~al.(2014)Garcin, Faltings, Donatsch, Alazzawi, Bruttin, and
  Huber]{swissinfo}
Florent Garcin, Boi Faltings, Olivier Donatsch, Ayar Alazzawi, Christophe
  Bruttin, and Amr Huber.
\newblock Offline and online evaluation of news recommender systems at
  swissinfo.ch.
\newblock In \emph{RecSys}, page 169–176, 2014.

\bibitem[Ghosh et~al.(2022)Ghosh, Ajay, Agrawal, and
  Levine]{adaptive-offline-rl}
Dibya Ghosh, Anurag Ajay, Pulkit Agrawal, and Sergey Levine.
\newblock Offline {RL} policies should be trained to be adaptive.
\newblock In \emph{ICML}, pages 7513--7530, 2022.

\bibitem[Gomez-Uribe and Hunt(2016)]{netflix}
Carlos~A. Gomez-Uribe and Neil Hunt.
\newblock The {Netflix} recommender system: Algorithms, business value, and
  innovation.
\newblock \emph{ACM Trans. Manage. Inf. Syst.}, 6\penalty0 (4), dec 2016.

\bibitem[GroupLens()]{movielens}
GroupLens.
\newblock {MovieLens} datasets.
\newblock URL \url{https://grouplens.org/datasets/movielens/}.

\bibitem[Gu et~al.(2022)Gu, Zhao, Chen, Li, Hao, and An]{merlion}
Pengjie Gu, Mengchen Zhao, Chen Chen, Dong Li, Jianye Hao, and Bo~An.
\newblock Learning pseudometric-based action representations for offline
  reinforcement learning.
\newblock In \emph{ICML}, pages 7902--7918, 2022.

\bibitem[Gulcehre et~al.(2020)Gulcehre, Wang, Novikov, Paine, Colmenarejo,
  Zo\l{}na, Agarwal, Merel, Mankowitz, Paduraru, Dulac-Arnold, Li, Norouzi,
  Hoffman, Heess, and Freitas]{rl-unplugged}
Caglar Gulcehre, Ziyu Wang, Alexander Novikov, Tom~Le Paine, Sergio~G\'{o}mez
  Colmenarejo, Konrad Zo\l{}na, Rishabh Agarwal, Josh Merel, Daniel Mankowitz,
  Cosmin Paduraru, Gabriel Dulac-Arnold, Jerry Li, Mohammad Norouzi, Matt
  Hoffman, Nicolas Heess, and Nando~de Freitas.
\newblock {RL} unplugged: A suite of benchmarks for offline reinforcement
  learning.
\newblock In \emph{NeurIPS}, 2020.

\bibitem[Hanna et~al.(2019)Hanna, Niekum, and Stone]{is-estimated-behavior}
Josiah Hanna, Scott Niekum, and Peter Stone.
\newblock Importance sampling policy evaluation with an estimated behavior
  policy.
\newblock In \emph{ICML}, pages 2605--2613, 2019.

\bibitem[Hidasi et~al.(2016)Hidasi, Karatzoglou, Baltrunas, and Tikk]{gru4rec}
Bal{\'{a}}zs Hidasi, Alexandros Karatzoglou, Linas Baltrunas, and Domonkos
  Tikk.
\newblock Session-based recommendations with recurrent neural networks.
\newblock In \emph{ICLR}, 2016.

\bibitem[Huang et~al.(2020)Huang, Oosterhuis, de~Rijke, and van Hoof]{sofa}
Jin Huang, Harrie Oosterhuis, Maarten de~Rijke, and Herke van Hoof.
\newblock Keeping dataset biases out of the simulation: A debiased simulator
  for reinforcement learning based recommender systems.
\newblock In \emph{RecSys}, page 190–199, 2020.

\bibitem[Hussein et~al.(2017)Hussein, Gaber, Elyan, and
  Jayne]{imitation-learning}
Ahmed Hussein, Mohamed~Medhat Gaber, Eyad Elyan, and Chrisina Jayne.
\newblock Imitation learning: A survey of learning methods.
\newblock \emph{ACM Comput. Surv.}, 50\penalty0 (2), apr 2017.

\bibitem[Instacart(2017)]{instacart}
Instacart.
\newblock Instacart market basket analysis, 2017.
\newblock URL
  \url{https://www.kaggle.com/c/instacart-market-basket-analysis/data}.

\bibitem[Jannach et~al.(2016)Jannach, Resnick, Tuzhilin, and
  Zanker]{beyond-matrix-completion}
Dietmar Jannach, Paul Resnick, Alexander Tuzhilin, and Markus Zanker.
\newblock Recommender systems — beyond matrix completion.
\newblock \emph{Commun. ACM}, 59\penalty0 (11):\penalty0 94–102, oct 2016.

\bibitem[Jeunen(2019)]{jeunen-offline-eval}
Olivier Jeunen.
\newblock Revisiting offline evaluation for implicit-feedback recommender
  systems.
\newblock In \emph{RecSys}, page 596–600, 2019.

\bibitem[Jeunen and Goethals(2021)]{optimizers-curse}
Olivier Jeunen and Bart Goethals.
\newblock Pessimistic reward models for off-policy learning in recommendation.
\newblock In \emph{RecSys}, page 63–74, 2021.

\bibitem[Ji et~al.(2021)Ji, Qin, Han, and Yang]{rl-ltv}
Luo Ji, Qi~Qin, Bingqing Han, and Hongxia Yang.
\newblock Reinforcement learning to optimize lifetime value in cold-start
  recommendation.
\newblock In \emph{CIKM}, page 782–791, 2021.

\bibitem[Ji et~al.(2020)Ji, Sun, Zhang, and Li]{data-leakage}
Yitong Ji, Aixin Sun, Jie Zhang, and Chenliang Li.
\newblock A critical study on data leakage in recommender system offline
  evaluation.
\newblock \emph{arXiv:2010.11060}, 2020.

\bibitem[Joachims et~al.(2017)Joachims, Swaminathan, and
  Schnabel]{ultr-joachims}
Thorsten Joachims, Adith Swaminathan, and Tobias Schnabel.
\newblock Unbiased learning-to-rank with biased feedback.
\newblock In \emph{WSDM}, page 781–789, 2017.

\bibitem[Kiyohara and Kawakami(2022)]{ofrl}
Haruka Kiyohara and Kosuke Kawakami.
\newblock {OFRL}: Designing an offline reinforcement learning and policy
  evaluation platform from practical perspectives.
\newblock In \emph{CONSEQUENCES+REVEAL@RecSys}, 2022.

\bibitem[Kostrikov et~al.(2021)Kostrikov, Nair, and Levine]{iql}
Ilya Kostrikov, Ashvin Nair, and Sergey Levine.
\newblock Offline reinforcement learning with implicit {Q}-learning.
\newblock \emph{arXiv:2110.06169}, 2021.

\bibitem[Krichene and Rendle(2020)]{sampled-metrics}
Walid Krichene and Steffen Rendle.
\newblock On sampled metrics for item recommendation.
\newblock In \emph{KDD}, page 1748–1757, 2020.

\bibitem[Kumar et~al.(2021)Kumar, Singh, Tian, Finn, and
  Levine]{workflow-offline-rl}
Aviral Kumar, Anikait Singh, Stephen Tian, Chelsea Finn, and Sergey Levine.
\newblock A workflow for offline model-free robotic reinforcement learning.
\newblock In \emph{CoRL}, 2021.

\bibitem[Last.fm()]{lastfm}
Last.fm.
\newblock URL \url{https://last.fm/api}.

\bibitem[Lee et~al.(2022)Lee, Hwang, Min, and Choo]{dongyoon}
Hojoon Lee, Dongyoon Hwang, Kyushik Min, and Jaegul Choo.
\newblock Towards validating long-term user feedbacks in interactive
  recommendation systems.
\newblock In \emph{SIGIR}, page 2607–2611, 2022.

\bibitem[Levine et~al.(2020)Levine, Kumar, Tucker, and Fu]{offline-tutorial}
Sergey Levine, Aviral Kumar, George Tucker, and Justin Fu.
\newblock Offline reinforcement learning: Tutorial, review, and perspectives on
  open problems.
\newblock \emph{arXiv:2005.01643}, 2020.

\bibitem[Ludewig et~al.(2019)Ludewig, Mauro, Latifi, and
  Jannach]{session-based-eval}
Malte Ludewig, Noemi Mauro, Sara Latifi, and Dietmar Jannach.
\newblock Performance comparison of neural and non-neural approaches to
  session-based recommendation.
\newblock In \emph{RecSys}, page 462–466, 2019.

\bibitem[McNee et~al.(2006)McNee, Riedl, and Konstan]{accurate-not-enough}
Sean~M. McNee, John Riedl, and Joseph~A. Konstan.
\newblock Being accurate is not enough: How accuracy metrics have hurt
  recommender systems.
\newblock In \emph{CHI}, page 1097–1101, 2006.

\bibitem[Oosterhuis and de~Rijke(2021)]{genspec}
Harrie Oosterhuis and Maarten~de de~Rijke.
\newblock Robust generalization and safe query-specializationin counterfactual
  learning to rank.
\newblock In \emph{WWW}, page 158–170, 2021.

\bibitem[{OpenAI} et~al.(2020){OpenAI}, Andrychowicz, Baker, Chociej,
  Jozefowicz, McGrew, Pachocki, Petron, Plappert, Powell, Ray, Schneider,
  Sidor, Tobin, Welinder, Weng, and Zaremba]{openai-rubiks-cube}
{OpenAI}, Marcin Andrychowicz, Bowen Baker, Maciek Chociej, Rafal Jozefowicz,
  Bob McGrew, Jakub Pachocki, Arthur Petron, Matthias Plappert, Glenn Powell,
  Alex Ray, Jonas Schneider, Szymon Sidor, Josh Tobin, Peter Welinder, Lilian
  Weng, and Wojciech Zaremba.
\newblock Learning dexterous in-hand manipulation.
\newblock \emph{The International Journal of Robotics Research}, 39\penalty0
  (1):\penalty0 3--20, 2020.

\bibitem[Precup et~al.(2000)Precup, Sutton, and Singh]{per-decision-IS}
Doina Precup, Richard~S. Sutton, and Satinder~P. Singh.
\newblock Eligibility traces for off-policy policy evaluation.
\newblock In \emph{ICML}, page 759–766, 2000.

\bibitem[Qin et~al.(2021)Qin, Gao, Zhang, Xu, Huang, Li, Zhang, and Yu]{neoRL}
Rongjun Qin, Songyi Gao, Xingyuan Zhang, Zhen Xu, Shengkai Huang, Zewen Li,
  Weinan Zhang, and Yang Yu.
\newblock {NeoRL}: A near real-world benchmark for offline reinforcement
  learning.
\newblock \emph{arXiv:2102.00714}, 2021.

\bibitem[Quadrana et~al.(2018)Quadrana, Cremonesi, and Jannach]{seq-aware-RS}
Massimo Quadrana, Paolo Cremonesi, and Dietmar Jannach.
\newblock Sequence-aware recommender systems.
\newblock \emph{ACM Comput. Surv.}, 51\penalty0 (4), jul 2018.

\bibitem[Reichlin et~al.(2022)Reichlin, Marchetti, Yin, Ghadirzadeh, and
  Kragic]{back-to-manifold}
Alfredo Reichlin, Giovanni~Luca Marchetti, Hang Yin, Ali Ghadirzadeh, and
  Danica Kragic.
\newblock Back to the manifold: Recovering from out-of-distribution states.
\newblock \emph{arXiv:2207.08673}, 2022.

\bibitem[Rendle et~al.(2019)Rendle, Zhang, and Koren]{tuning-baselines}
Steffen Rendle, Li~Zhang, and Yehuda Koren.
\newblock On the difficulty of evaluating baselines: A study on recommender
  systems.
\newblock \emph{arXiv:1905.01395}, 2019.

\bibitem[RetailRocket(2016)]{retailrocket}
RetailRocket.
\newblock {RetailRocket} recommender system dataset, 2016.
\newblock URL
  \url{https://www.kaggle.com/datasets/retailrocket/ecommerce-dataset}.

\bibitem[Shi et~al.(2019)Shi, Yu, Da, Chen, and Zeng]{taobao}
Jing{-}Cheng Shi, Yang Yu, Qing Da, Shi{-}Yong Chen, and Anxiang Zeng.
\newblock Virtual-taobao: Virtualizing real-world online retail environment for
  reinforcement learning.
\newblock In \emph{AAAI}, pages 4902--4909, 2019.

\bibitem[Sun(2022)]{fresh-look-RS}
Aixin Sun.
\newblock From counter-intuitive observations to a fresh look at recommender
  system.
\newblock \emph{arXiv:2210.04149}, 2022.

\bibitem[Sun et~al.(2020)Sun, Yu, Fang, Yang, Qu, Zhang, and
  Geng]{benchmarking-reco}
Zhu Sun, Di~Yu, Hui Fang, Jie Yang, Xinghua Qu, Jie Zhang, and Cong Geng.
\newblock Are we evaluating rigorously? {Benchmarking} recommendation for
  reproducible evaluation and fair comparison.
\newblock In \emph{RecSys}, page 23–32, 2020.

\bibitem[Sutton and Barto(2018)]{sutton-barto}
Richard Sutton and Andrew Barto.
\newblock \emph{Reinforcement Learning: An Introduction}.
\newblock MIT Press, 2018.

\bibitem[Swaminathan and Joachims(2015)]{crm-joachims}
Adith Swaminathan and Thorsten Joachims.
\newblock Batch learning from logged bandit feedback through counterfactual
  risk minimization.
\newblock \emph{Journal of Machine Learning Research}, 16\penalty0
  (52):\penalty0 1731--1755, 2015.

\bibitem[Tobin et~al.(2017)Tobin, Fong, Ray, Schneider, Zaremba, and
  Abbeel]{domain-randomization}
Josh Tobin, Rachel Fong, Alex Ray, Jonas Schneider, Wojciech Zaremba, and
  Pieter Abbeel.
\newblock Domain randomization for transferring deep neural networks from
  simulation to the real world.
\newblock In \emph{2017 IEEE/RSJ International Conference on Intelligent Robots
  and Systems (IROS)}, pages 23--30, 2017.
\newblock \doi{10.1109/IROS.2017.8202133}.

\bibitem[van Hasselt et~al.(2018)van Hasselt, Doron, Strub, Hessel, Sonnerat,
  and Modayil]{deadly-triad}
Hado van Hasselt, Yotam Doron, Florian Strub, Matteo Hessel, Nicolas Sonnerat,
  and Joseph Modayil.
\newblock Deep reinforcement learning and the deadly triad.
\newblock \emph{arXiv:1812.02648}, 2018.

\bibitem[Zhao et~al.(2022)Zhao, Lin, Feng, Wang, and Wen]{revisiting-study}
Wayne~Xin Zhao, Zihan Lin, Zhichao Feng, Pengfei Wang, and Ji-Rong Wen.
\newblock A revisiting study of appropriate offline evaluation for top-n
  recommendation algorithms.
\newblock \emph{ACM Trans. Inf. Syst.}, jun 2022.

\end{thebibliography}
\end{document}